\documentclass[letter,preprintnumbers,nofootinbib,superscriptaddress,twocolumn]{revtex4}
\usepackage{amssymb}
\usepackage{graphicx}
\usepackage{dcolumn}
\usepackage{bm}

\begin{document}

\title{Peak-Dip-Hump from Holographic Superconductivity}
\author{Jiunn-Wei Chen}
\email{jwc@phys.ntu.edu.tw}
\affiliation{Department of Physics and Center for Theoretical Sciences}
\author{Ying-Jer Kao}
\email{yjkao@phys.ntu.edu.tw}
\affiliation{Department of Physics and Center for Theoretical Sciences}
\author{Wen-Yu Wen}
\email{steve.wen@gmail.com}
\affiliation{Department of Physics and Center for Theoretical Sciences}
\affiliation{Leung Center for Cosmology and Particle Astrophysics\\
National Taiwan University, Taipei 106, Taiwan}

\begin{abstract}
We study the fermionic spectral function in a holographic superconductor
model. At zero temperature, the black hole has zero horizon and hence the
entropy of the system is zero after the back reaction of the condensate is
taken into account. We find the system exhibits the famous peak-dip-hump
lineshape with a sharp low-energy peak followed by a dip then a hump at
higher energies. This feature is widely observed in the spectrum of several
high-$T_{c}$ \ superconductors. We also find a linear relation between the
gap in the fermionic spectrum and the condensate, indicating the condensate
is formed by fermion pairing.
\end{abstract}

\pacs{}
\maketitle





\textit{Introduction}--The proper description of the physics of strongly
interacting fermions has long been the major issue for our understanding of
the quark-gluon interactions and condensed matter systems near quantum
criticality. Recent development of the holographic correspondence between a
gravitational theory and a quantum field theory, which first emerged under
the anti-de-Sitter space/conformal field theory (AdS/CFT) correspondence 
\cite{Maldacena:1997re,Gubser:1998bc,Witten:1998qj}, has provided a useful
framework to describe systems with strong interactions (see e.g. \cite%
{Policastro:2001yc,Herzog:2006gh,Liu:2006ug,Gubser:2006bz,Herzog:2007ij,Hartnoll:2007ih,Hartnoll:2007ip,Hartnoll:2008hs}%
).

Recently, a gravitational model of hairy black holes \cite%
{Gubser:2005ih,Gubser:2008px}, where the Abelian symmetry of Higgs is
spontaneously broken below some critical temperature, has been used to model
high-$T_{c}$ superconductivity (HTSC) \cite{Hartnoll:2008vx}. On the fermion
side, it has been shown that Fermi liquids emerge by deforming away from the
quantum critical point by increasing the fermion density \cite{CZS}. Also, a
theory describing non-Fermi liquids, which describes the normal state of the
high-$T_{c}$ superconducting cuperates and metals near quantum critical
points, has been proposed \cite{Liu:2009dm}. The response functions of
composite fermionic operators in a class of strongly interacting quantum
field theories at finite density are computed using the AdS/CFT
correspondence \cite{Lee:2008xf,Liu:2009dm}. In particular, in Ref. \cite%
{Liu:2009dm} the Dirac equation is probed in the background of an extremal
AdS$_{4}$ charged black hole which possesses a finite horizon area and thus
finite entropy at temperature $T=0$. Analysis of the fermionic spectral
function gives gapless fermionic excitations at discrete shells in momentum
space, signaling the existence of a Fermi surface, and the spectral weight
exhibits novel properties, for example, particle-hole asymmetry.

In this letter, we study the fermionic excitations of a holographic
superconductor. This is motivated by the great importance to understand how
fermionic quasiparticle excitation is affected by the condensate, as this
may hold the key to the mystery of HTSC. As an intrinsic strong coupling
system, HTSC has posted great theoretical difficulties since its discovery.
With the AdS/CFT correspondence, where a strongly coupled field theory can
be mapped to a weakly curved gravity theory, nonperturbative computations in
quantum field theory turns into solving classical Einstein gravity. At $T=0$%
, the symmetry broken phase has degenerate vaccua. However, since only one
state is accessible, it is more reasonable to choose a gravitational model
with zero horizon at $T=0$. This requirement can be achieved by including
the full back reaction \cite{Horowitz:2009ij} of the condensate to the
gravity geometry \cite{Hartnoll:2008kx}. By applying the fermion spectral
function technique developed in \cite{Iqbal:2009fd}, we find that the system
exhibits the famous peak-dip-hump (PDH) lineshape with a sharp low-energy
peak followed by a dip then a hump at higher energies. This feature is
widely observed in the spectrum of several HTSC, for example, in the angle
resolved photoemission spectroscopy (ARPES) data near $(\pi ,0)$ point of
the Brilluoin zone in Bi$_{2}$Sr$_{2}$CaCu$_{2}$O$_{8}$(Bi2212) \cite%
{Norman:1997, Shen:2003}, and in tunneling experiments \cite{Tallon:2001}.
This feature has been attributed to the many-body effects of the fermionic
quasiparticle coupling to a bosonic collective mode, such as the $(\pi ,\pi )
$ resonance mode in inelastic neutron scattering \cite{Neutron}, and it is
argued to be related to the microscopic pairing interaction. Understanding
the physics of this PDH lineshape feature is therefore very important for
the construction of a microscopic mechanism for HTSC. Furthermore, by
assigning the fermion charge half that of the condensate, we find a linear
relation between the gap in the fermionic spectrum and the condensate. This
result indicates that the condensate is related to the fermion pair in our
theory. Since we use fermionic degrees of freedom as a probe to the
condensate, this might provide us a pairing mechanism without a "glue" for
superconductivity.


\textit{Gravitational Model of the Superconductor}--Inspired by the AdS/CFT
correspondence, we conjecture that there is a strongly interacting CFT dual
to the gravitational model that we will describe below. It is assumed that
even though we do not know what exactly the underlying field theory is, we
can still study its properties from the gravitational side, using the
AdS/CFT machinery.

We start with the following action with a vector field $A_{\mu }$ and a
scalar field $\psi $ coupled to a AdS$_{4}$ background \cite{Hartnoll:2008kx}%
, 
\begin{eqnarray}
S &=&\frac{1}{2\kappa ^{2}}\int {d^{4}x}\sqrt{-g}\big\{R+\frac{6}{L^{2}} 
\nonumber \\
&&-\frac{1}{4}F^{2}-|\partial \psi -iqA\psi |^{2}+m^{2}{|\psi |^{2}}\big\},
\label{S}
\end{eqnarray}%
where the gravitational coupling $\kappa ^{2}=8\pi G_{N}$, $m$ and $q$ are
the mass and charge of the scalar field. We will use the units in which the
radius of curvature $L=1$. A fully back-reacted ansatz for the metric is 
\begin{eqnarray}
&&ds^{2}=-g(r)e^{-\chi (r)}dt^{2}+\frac{dr^{2}}{g(r)}+r^{2}(dx^{2}+dy^{2}), 
\nonumber \\
&&\mathrm{with}\quad A=\phi (r)dt,\qquad \psi =\psi (r).  \label{Metric}
\end{eqnarray}%
Its zero temperature limit has been studied in detail in \cite%
{Horowitz:2009ij}. Here, we will focus on the $m^{2}=0$ case as the metric
is non-sigular near the horizon ($r=0$): 
\begin{eqnarray}
&&\phi =r^{2+\alpha },\qquad \psi =\psi _{0}-\psi _{1}r^{2(1+\alpha )}, 
\nonumber \\
&&\chi =\chi _{0}-\chi _{1}r^{2(1+\alpha )},\quad
g=r^{2}(1-g_{1}r^{2(1+\alpha )}),  \label{horizon}
\end{eqnarray}%
where $\psi _{0},\psi _{1},\chi _{1}$ and $g_{1}$ are all functions of $%
\alpha $ and $\chi _{0}$, and their explicit forms are omitted here. Note
that the metric becomes an AdS$_{4}$ near the horizon. At the infinite
boundary ($r\rightarrow \infty $), 
\begin{eqnarray}
&&\phi =\mu -\frac{\rho }{r}+\cdots ,\quad \psi =J+\frac{\left\langle 
\mathcal{O}_{B}\right\rangle }{r^{3}}+\cdots ,  \nonumber  \label{boundary}
\\
&&\chi \rightarrow 1,\quad g\rightarrow r^{2},
\end{eqnarray}%
where $\mu $ and $\rho $ have the holographic interpretation as chemical
potential and density, while $J$ and $\left\langle \mathcal{O}%
_{B}\right\rangle $ are the source and the condensate of the bosonic
operator $O_{B}$ which is dual to $\psi $\ and living on the boundary. We
use the shooting method to find the values of $\alpha $\ and $\chi _{0}$\
that gives $J=0$\ (sourceless condition) and a chosen value of $\rho $\ (or $%
\mu $) at the boundary. In the above choice of boundary condition on $\psi $
we have set the conformal dimension of $\mathcal{O}_{B}$ as $\Delta _{B}=3$,
while $g$ and $\chi $ asymptote to an AdS$_{4}$ at $r\rightarrow \infty $ 
\footnote{%
In the calculation, $\chi $ could approach some nonzero constant at $%
r\rightarrow \infty $, implying a gravitational redshift. One can rescale $%
t\rightarrow e^{-\chi /2}t$ and $\phi \rightarrow e^{\chi /2}\phi $ to the
Einstein frame where the speed light is normalized to $1$.}.

Keeping $\rho $ constant, we obtain the following numerical results by
fitting,

\begin{eqnarray}
\mu &=&2.62\sqrt{\frac{\rho }{q}}(1-e^{-q/0.8}),  \nonumber \\
\left\langle \mathcal{O}_{B}\right\rangle &=&0.025\mu ^{3}q^{2}\tanh
(q-1.05).  \label{condensate}
\end{eqnarray}%
The $\mu $ dependence follows directly from dimensional analysis. For
constant $\rho $, $\mu $ is monotonically decreasing in $q$ while $%
\left\langle \mathcal{O}_{B}\right\rangle $ is monotonically increasing in $%
q $. From this we know that the interaction is attractive. Because it is
increasingly difficult to approach small $q$ using the shooting method, our
fit cannot determine precisely the critical value of $q$ below which the
condensate vanishes nor the order of this phase transition.

\textit{Fermionic Spectral Function}-- The fermionic spectral function is a
powerful tool to probe the properties of the superconductor. We study the
two point correlator (a retarded Green's function $G_{R}$) of the fermionic
operator $\mathcal{O}_{F}$, and we will focus on the spin averaged spectral
function, which is defined as the imaginary part of the trace of $G_{R}$ (in
the space of the energy $\omega $ and the momentum $k$) 
\begin{equation}
A(\omega ,k)\equiv \text{Im}(\text{Tr}G_{R}).
\end{equation}%
Note that $G_{R}$ can either be computed using the CFT or through the
fermionic field $\Psi $ which lives in the bulk but couples to $\mathcal{O}%
_{F}$ at the boundary. In the second approach, $\mathcal{O}_{F}$ can turn
into $\Psi $. Hence, the two point function of $\mathcal{O}_{F}$ is related
to the two point function of $\Psi $ at the boundary. One can show that
summing the corresponding momentum modes in the $r$ direction gives $%
A(\omega ,k)=2$ as $|\omega |\rightarrow \infty $. For comparison, a free
fermion spectral function on the boundary of a 4D flat space is\textbf{\ }%
\begin{equation}
A(\omega ,k)=2\left\vert \omega \right\vert /\sqrt{\omega ^{2}-k^{2}}\theta
\left( \left\vert \omega \right\vert -\left\vert k\right\vert \right) 
\mathbf{.}  \label{free}
\end{equation}

In our case, the information needed to compute $G_{R}$ can be obtained by
solving the Dirac equation. With $\Psi $ minimally coupled to the gauge
field $A_{\mu }$\ in the gravitational background of Eq.(\ref{Metric}), the
Dirac equation can be recast into the flow equation \cite{Liu:2009dm} 
\begin{eqnarray}
&&\sqrt{\frac{g_{ii}}{g_{rr}}}\partial _{r}\xi _{\pm }=-2m_{F}\sqrt{g_{ii}}%
\xi _{\pm }\mp (k\mp u(r))\pm (k\pm u(r))\xi _{\pm }^{2},  \nonumber \\
&&\text{with}\quad u(r)=\sqrt{\frac{g_{ii}}{-g_{tt}}}(\omega +q_{e}\phi (r)),
\label{flow}
\end{eqnarray}%
where $q_{e}$ is the fermion charge. The mass $m_{F}$ of $\Psi $ is related
to $\Delta _{F}$, the mass dimension of $\mathcal{O}_{F}$, as $\Delta
_{F}=m_{F}+3/2$. Because of the strong interaction nature of the problem and
the lack of supersymmetry in our background, it is not clear how the charge
and mass of the scalar field is related to those of the fermion field.
Nevertheless, we will focus on a possible scenario: $q=2q_{e}$ and $%
m_{F}=m=0 $. Thus, the scalar field has the same charge as a fermion pair,
and the dimension of the bosonic operator $\mathcal{O}_{B}$ is just twice of
the fermionic operator $\mathcal{O}_{F}$ ($\Delta _{B}=2\Delta _{F}$). With $%
m_{F}=0$, we have%
\begin{equation}
\text{Tr}G_{R}=\left( \xi _{+}+\xi _{-}\right) |_{r\rightarrow \infty }.
\label{Tr}
\end{equation}%
\begin{figure}[tbp]
\label{fig3} \includegraphics[width=0.45\textwidth]{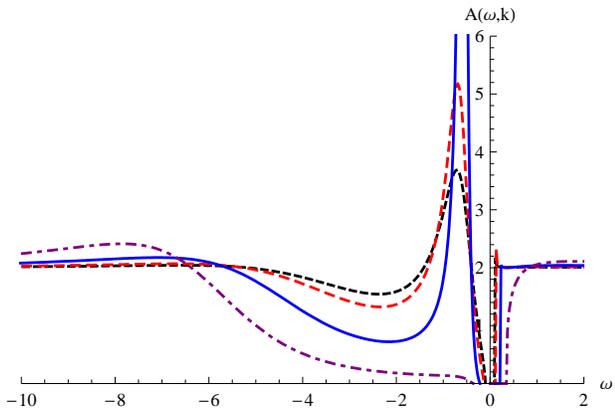}
\caption{ A typical fermionic spectral function with the peak-dip-hump
structure for $k=0.7$ (black dotted) $0.9$ (red dashed) $1.5$ (blue line) $%
2.5$ (purple dotdashed). A sharp peak develops when $k$ increases and the
peak disappears above the Fermi momentum $k_{F}$. Here, $\protect\rho =10$
and $q=1.7$. }
\end{figure}

\begin{figure}[tbp]
\label{fig2} \includegraphics[width=0.45\textwidth]{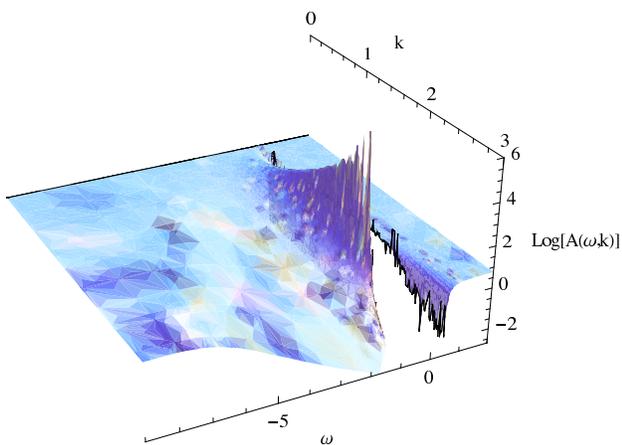}
\caption{3D plot of the fermionic spectral function with the peak-dip-hump
structure. Here, $\protect\rho =10$ and $q=1.7$. }
\end{figure}

\textit{Results and Discussions}-- Figure 1 shows the spectral function $%
A(\omega ,k)$ of $(q,\rho )=(1.7,10)$ for sampled values of $k$. $A(\omega
,k)$ is even in $k$ as $\xi _{+}\left( -k\right) =\xi _{-}(k)$ in Eq.(\ref%
{flow}) for $m_{F}=0$. So we only need to focus on positive $k$. We see at
large $\left\vert \omega \right\vert $ $A(\omega ,k)$ asymptotes to 2, as
mentioned above. There is an $A(\omega ,k)=0$ region bounded by $\left\vert
\omega \right\vert <v_{0}\left\vert k\right\vert $, where $v_{0}$ is $k$
independent and $v_{0}\leq 1$. Outside this region, there appears a
peak-dip-hump structure in the negative $\omega $ (hole excitation) side
when $k$ is below some critical value $k_{F}$. As $k $ increases (but still
below $k_{F}$), the peak gets sharper and taller. Furthermore, both the peak
and dip move toward smaller $\left\vert \omega \right\vert $ while the hump
moves toward bigger $\left\vert \omega \right\vert $. The corresponding 3D
plot of log$A(\omega ,k)$ is shown in Fig. 2. It can be seen clearly that
the peak becomes sharpest and tallest close to $k_{F}$ and disappears just
above $k_{F}$. It is encouraging that this behavior is very similar to what
is observed in the superconducting phase of several HTSC materials.

We identify $k_{F}$ as the Fermi momentum, because closer to the Fermi
surface a hole excitation has fewer neighboring particles to annihilate
with, so that its peak is sharper. We also identify the excitation energy of
the sharpest peak as the gap $\Delta $. It is consistent with the
conventional definition of the lightest quasiparticle excitation energy.
Numerically, for a wide range of \ $q$ and $\mu $ (or $\rho $) values, we
find that the gap (in the fermionic spectrum) is related to the bosonic
condensate as 
\begin{equation}
\Delta \simeq \frac{0.6\left\langle \mathcal{O}_{B}\right\rangle }{\rho }.
\label{gap}
\end{equation}%
This $\Delta \propto \left\langle \mathcal{O}_{B}\right\rangle $ relation is
also found in the BCS theory. It shows that the dynamics of the fermion and
boson fields in the holographic theory are intimately related.

It is instructive to show parametrically how the above relation could emerge
out of the bulk theory. In the bulk, the term $|qA\psi |^{2}$ in Eq.(\ref{S}%
) gives the gauge field a mass $q^{2}|\psi |^{2}$ (Higgs mechanism). The
exchange of one massive gauge boson gives a localized potential between two
fermions which can then be written as an effective four fermion contact
interaction with coupling $C\simeq 1/|\psi |^{2}$. This interaction can
contribute to the fermion self energy diagram with the fermion mass (i.e.
the gap $\Delta $) determined self-consistently. This gap equation is solved
when the fermion loop diagram $I=1/C$. Since the fermion loop diagram has
two fermion propagators and a four momentum integral, dimensional analysis
gives $I\propto $ $\Delta ^{2}$.\ Thus, we have $\Delta \propto |\psi
|\propto \left\langle \mathcal{O}_{B}\right\rangle $.

Now we reach an interesting conclusion from the above discussions. From the
analogy to the BCS theory, the bosonic condensate is the consequence of the
fermion pairing. On the other hand, the condensate is also causing the
pairing, because the condensate gives the gauge boson mass through the Higgs
mechanism which is important for the fermion pairing.

To understand the $A(\omega ,k)=0$ region bounded by $\left\vert \omega
\right\vert <v_{0}\left\vert k\right\vert $, it is easier to go to the $%
q\rightarrow \infty $ limit. By rescaling $A\rightarrow A/q$ and $\psi
\rightarrow \psi /q$ in Eq.(\ref{S}), $A$ and $\psi $ decouple from the
Einstein equation; their back reaction can be neglected so the background is
just AdS$_{4}$ \cite{Hartnoll:2008vx}. In this \textquotedblleft probe
limit,\textquotedblright\ the bulk fermion satisfies the 4D Klein-Gordon
equation 
\begin{equation}
\left[ -\nabla _{r}^{2}+k^{2}-(\omega +q_{e}\phi (r))^{2}\right] \psi =0.
\label{KG}
\end{equation}%
This is similar to a Shrodinger equation with a potential $V(r)=-(\omega
+q_{e}\phi (r))^{2}$ and energy $-k^{2}$. $\phi (r)$ is zero at $r=0$, and
it increases smoothly and monotonically to $\mu $ at $r\rightarrow \infty $.
One can show that when $\left\vert \omega \right\vert <\left\vert
k\right\vert $, $\psi $ is a bound state in the $r$-direction. This implies
the bulk fermion propagation is not on-shell and hence Im$(G_{R})$ and $%
A(\omega ,k)$ vanish. Away from the probe limit, the back reaction modifies
the metric near $r=0$ but not at large $r$. Near $r=0$, the light velocity
factor\ $v(r)\equiv \sqrt{-g_{ii}/g_{tt}}$ starts to deviate from unity and
decreases with decreasing $q$. The potential of the Schrodinger equation is
then modified to $V(r)=-v^{2}(r)(\omega +q_{e}\phi (r))^{2}$, which makes
the $A(\omega ,k)=0$ region to be bounded by $\left\vert \omega \right\vert
<v_{0}\left\vert k\right\vert $ with $v_{0}=v(0)$ ($v_{0}=1$ in the probe
limit). In Figs. 1 and 2, $v_{0}=0.15$\ for $q=1.7$.

Now we discuss the physics of the hump. The hump has the dispersion relation 
\begin{equation}
\omega _{hump}\simeq -q_{e}\mu -k,
\end{equation}%
and its width is $\sim \mathcal{O}(q_{e}\mu )$. The dispersion comes from
the fermion gauge coupling to $A_{t}$ on the \ boundary, which gives an
energy shift $\omega \rightarrow \omega +q_{e}\mu $ to the dispersion $%
\omega =-k$ of massless hole fermions. This is a mode of fermion propagation
parallel to the boundary. More specifically, near the boundary, the bulk
fermion satisfies Eq. (\ref{KG}). Thus, if $\phi (r)$ were a constant $\mu $%
, the spectral function would be similar to Eq. (\ref{free}) with a sharp
peak at $\omega =-q_{e}\mu -k$. In reality, $\phi (r)$ almost remains
constant except at small $r$. This curvature introduces a smearing effect of
the sharp peak and sets the scale of the hump width. The physics picture is
that the gradient of $\phi (r)$ shows that the electric field localizes near 
$r=0$. This is just in analogy with the Meissner effect which is a
consequence of the Higgs mechanism. The photon mass $q|\psi |$ sets the
scale of the hump width. Numerically, this scale is $\sim q\mu $ ($\mu $ is
the only dynamical scale in the problem) so the width is as large as $\omega
_{hump}$ at smaller $k$. This is indeed what we have seen in Fig. 1.

In the above example, we have seen that while the size of $\rho $ does not
affect the physical properties we discussed above, the size of $q$ could
cause more significant changes by controlling the geometry. Another example
is how the peak position changes with respect to $k$. In Fig. 1, the peak
moves toward smaller $\left\vert \omega \right\vert $ with increasing $k$.
In general, this is the case for $q\lesssim 2.6$. For $q\gtrsim 2.6$, the
peak moves toward bigger $\left\vert \omega \right\vert $ when $k$
increases. This might imply that the system could go through a quantum phase
transition by changing $q$, which requires further studies.


In conclusion, we have studied the spectrum of fermionic excitations in a
gravitational model which has taken account of the back reaction of the
condensate. The system exhibits the PDH lineshape in the fermionic spectral
function which shows great similarity to that in the ARPES experiments of
HTSC. In addition, we have found a linear relation between the gap in the
fermionic spectrum and the condensate, indicating the condensate is formed
by fermion pairing. This is highly non-trivial, and it opens possibilities
to study new microscopic mechanisms for superconductivity without explicit
pairing interactions.

\begin{acknowledgments}
We thank Chyh-Hong Chern, Pei-Ming Ho, Feng-Li Lin, and Chen-Pin Yeh for
useful discussions and Udit Raha for careful reading of the manuscript. This
work was supported in part by the National Science Council and National
Center for Theoretical Sciences of R.O.C. under grants NSC
96-2112-M-002-019-MY3 (JWC) 97-2628-M-002 -011 -MY3 (YJK),
97-2112-M-002-015-MY3 (WYW).
\end{acknowledgments}


\bibliographystyle{plain}
\bibliography{apssamp}

\end{document}